\documentclass[twocolumn,final,prd,superscriptaddress,showpacs,floatfix,noshowpacs,
,nofootinbib]{revtex4}

\usepackage[final]{epsfig}
\usepackage{bm}
\usepackage{mathtext}
\usepackage{amssymb, amsmath}
\usepackage{color}

\def\Xint#1{\mathchoice
{\XXint\displaystyle\textstyle{#1}}%
{\XXint\textstyle\scriptstyle{#1}}%
{\XXint\scriptstyle\scriptscriptstyle{#1}}%
{\XXint\scriptscriptstyle\scriptscriptstyle{#1}}%
\!\int}
\def\XXint#1#2#3{{\setbox0=\hbox{$#1{#2#3}{\int}$ }
\vcenter{\hbox{$#2#3$ }}\kern-.6\wd0}}

\def\dashint{\Xint-}

\begin{document}

\title{Photon-photon dispersion of TeV gamma rays
and its role for photon-ALP conversion}

\author{Alexandra Dobrynina}
\affiliation{Max-Planck-Institut f\"ur Physik (Werner-Heisenberg-Institut),
  F\"ohringer Ring 6, 80805 M\"unchen, Germany}
\affiliation{P.~G.~Demidov Yaroslavl State University,
  Sovietskaya 14, 150000 Yaroslavl, Russia}

\author{Alexander Kartavtsev}
\affiliation{Max-Planck-Institut f\"ur Physik (Werner-Heisenberg-Institut),
  F\"ohringer Ring 6, 80805 M\"unchen, Germany}

\author{Georg Raffelt}
\affiliation{Max-Planck-Institut f\"ur Physik (Werner-Heisenberg-Institut),
  F\"ohringer Ring 6, 80805 M\"unchen, Germany}

\date{15 December 2014, revised 21 February 2015, corrected 29 April 2015 and 8 May 2017}

\begin{abstract}
The propagation of TeV gamma rays can be strongly modified by $B$-field
induced conversion to axionlike particles (ALPs). We show that, at such
high energies, photon dispersion is dominated by background photons---the
only example where photon-photon dispersion is of practical relevance. We
determine the refractive index for all energies and find that, for fixed
energy density, background photons below the pair-production threshold
dominate. The cosmic microwave background alone provides an ``effective
photon mass'' of $m_\gamma^2=-(1.01~{\rm neV}\times\omega/{\rm TeV})^2$ for
$\omega\alt1000$~TeV. The extragalactic background light is subdominant,
but local radiation fields in the galaxy or the source regions provide
significant contributions. Photon-photon dispersion is small enough to
leave typical scenarios of photon-ALP oscillations unscathed, but big
enough to worry about it case by case.
\end{abstract}


\maketitle

\section{Introduction}

Astronomy with TeV gamma rays has opened a new window to the universe,
allowing us to study a plethora of fantastic sources of very high-energy
photons~\cite{Aharonian:2004yt, Aharonian:2013, Hinton:2009zz, Lorenz:2012nw,
Paneque:2012kv}. In addition to the sources themselves, we can study
intervening phenomena. In particular, the radiation emitted by all stars, the
extragalactic background light (EBL), absorbs photons by $\gamma\gamma\to
e^+e^-$. As a result, the \hbox{TeV $\gamma$-ray} horizon is only some
100~Mpc and the observed source spectra are strongly modified. One can use
this effect to explore the EBL which is hard to measure directly
\cite{Stecker:1992wi, Dwek:2012nb, Biteau:2015xpa}. More fundamentally, the
fast time structure of certain sources allows one to constrain novel
dispersion effects, for example by Lorentz invariance violation
\cite{Stecker:2001vb, Liberati:2009pf, Liberati:2013xla}.

We are here concerned with another effect at the low-energy frontier of
elementary particle physics \cite{Kuster:2008zz, Jaeckel:2010ni,
Hewett:2012ns, Agashe:2014kda}, the conversion of photons into axionlike
particles (ALPs) in large-scale magnetic fields \cite{Sikivie:1983ip,
Raffelt:1987im}, enabled by the two-photon vertex of these hypothetical
low-mass bosons. The conversion $\gamma\to a$ modifies the source spectra.
The conversion and subsequent back conversion $\gamma\to a\to\gamma$ allows
TeV gamma rays to ``propagate in disguise'' and evade absorption by $e^+e^-$
pair production \cite{Csaki:2003ef, Mirizzi:2007hr, Hooper:2007bq, De
Angelis:2007yu, De Angelis:2007dy, Hochmuth:2007hk, Simet:2007sa,
DeAngelis:2008sk, Fairbairn:2009zi, SanchezConde:2009wu, Bassan:2009gy,
Mirizzi:2009aj, Dominguez:2011xy, DeAngelis:2011id, Tavecchio:2012um,
Wouters:2012qd, Horns:2012kw, Mena:2013baa, Wouters:2013eka, Meyer:2013pny,
Meyer:2014epa, Tavecchio:2014yoa, Reesman:2014ova, Harris:2014nza,
Meyer:2014gta, Galanti:2015rda}. This effect is a possible explanation of the cosmic
transparency problem, i.e., TeV gamma rays seem to travel further than
allowed by typical EBL estimates. At the very least, this effect represents a
systematic uncertainty when probing the EBL with TeV gamma rays.

Photon and ALP propagation and conversion is most easily studied in analogy
to neutrino flavor oscillations \cite{Raffelt:1987im, Dasgupta:2010ck}.  A
wave of frequency $\omega$ and amplitude $A$ evolves in the $x$ direction
according to $-i\partial_x A=n_{\rm refr}\omega\,A$, where $n_{\rm refr}$ is
the refractive index which gives us the wave number by $k=n_{\rm
refr}\omega$.  We write $n_{\rm refr} = 1+\chi+i\kappa$ and assume
$|\chi+i\kappa|\ll1$. The real part $\chi$ describes dispersion and the
imaginary part $\kappa$ absorption. $A$ has three components, the photon
amplitude $A_\perp$ with polarization perpendicular to the transverse
$B$-field, $A_\parallel$ parallel to it, and the ALP amplitude $a$, i.e.,
$A=(A_\perp, A_\parallel, a)$, and $\chi$ and $\kappa$ are now 3$\times$3
matrices. The off-diagonal $\chi$ elements cause oscillations between
different $A$-components such as the Faraday effect, where electrons in the
longitudinal $B$-field instigate a rotation of the plane of polarization.

ALPs interact with photons by ${\cal L}_{a\gamma}=g_{a\gamma}{\bf E}\cdot{\bf
B}\,a$ in terms of the electric, magnetic, and ALP fields whereas
$g_{a\gamma}$ is a coupling constant of dimension inverse energy. An external
transverse magnetic field $B_{\rm T}$ couples $A_\parallel$ with $a$ and
provides an off-diagonal refractive index $\chi_{a\gamma}=g_{a\gamma}B_{\rm
T}/2\omega$ which leads to ALP-photon oscillations. (We always use natural
units with $\hbar=c=k_{\rm B}=1$.) The ALP dispersion relation is
$\omega^2-k^2=m_a^2$, providing the refractive index
$\chi_a=-m_a^2/2\omega^2$. An analogous expression pertains to photons where
the plasma frequency $\omega_{\rm pl}^2=4\pi\alpha n_e/m_e$ is the effective
photon mass.

More important for TeV $\gamma$-ray dispersion is the $B$-field itself due to
an effective photon-photon interaction mediated by virtual $e^+e^-$ pairs. At
low energies, it is described by the Euler-Heisenberg Lagrangian ${\cal
L}_{\gamma\gamma}=(2\alpha^2/45 m_e^4)\, [({\bf E}^2-{\bf B}^2)^2+7({\bf
E}\cdot{\bf B})^2]$, which however also pertains to background
photons. The overall electromagnetic (EM) energy density $\rho_{\rm
EM}=\frac{1}{2}\langle E^2+B^2\rangle$ produces \cite{Tarrach:1983cb,
Barton:1989dq, Latorre:1994cv, Kong:1998ic, Thoma:2000fd, Dittrich:2000zu}
\begin{equation}\label{eq:chiEM}
\chi_{\rm EM}=\frac{44\alpha^2}{135}\,\frac{\rho_{\rm EM}}{m_e^4}\,,
\end{equation}
implying spacelike dispersion $\omega^2-k^2=-2\chi_{\rm EM}\omega^2$, i.e.,
a ``negative effective mass-squared.''

Large-scale fields or nonisotropic background photons imply further
geometrical factors depending on direction of motion and
polarization. If the EM background is a homogeneous $B$-field, the
dispersion of $A_\parallel$ receives a factor\footnote{In
    an earlier version of this Eprint, published in PRD, we had given
    $14/11$ and $8/11$ for these coefficients, respectively, a factor
    2/3 smaller. We had mistakenly included
    $\langle\sin^2\theta\rangle=2/3$ in these expressions. We thank
    M.~Roncadelli for pointing out this error. Notice that for an
    isotropic gas of unpolarized test photons, averaging over directions
    leads to $\langle\sin^2\theta\rangle=2/3$ and averaging over polarizations
    leads to $(1/2)\times(21/11 + 12/11)=3/2$, i.e., their average
    dispersion in a $B$-field
    is indeed given by Eq.~(\ref{eq:chiEM}).}
  $(21/11)\,\sin^2\theta$, whereas $A_\perp$ a factor
  $(12/11)\,\sin^2\theta$ \cite{Dittrich:2000zu, Toll:1952rq,
    Adler:1971wn, Tsai:1975iz, Karbstein:2013ufa}. Here, $\theta$ is
  the angle between the photon and $B$-field directions, i.e., only
  the transverse field strength $B_{\rm T}$ enters. These results
  correspond to what has been used in studies of TeV $\gamma$-ray
  propagation.

A minimal EM energy density everywhere is $\rho_{\rm CMB} =
(\pi^2/15)\,T^4=0.261~{\rm eV}~{\rm cm}^{-3}$ provided by the cosmic
microwave background (CMB), where we have used $T=2.726~{\rm K}$,
leading to
\begin{equation}
\label{chiCMB}
\chi_{\rm CMB}=0.511\times10^{-42}\,.
\end{equation}
A typical galactic $B$-field of $1~\mu{\rm G}$ corresponds to $\rho_{\rm EM}
= 0.0248~{\rm eV}~{\rm cm}^{-3}$, so the CMB dominates by a factor of ten.
Depending on the environment, dispersion of TeV gamma rays involves the CMB
and possible larger local radiation densities. This insight is our main
point.

\section{Photon-ALP oscillations}

To develop a sense of the quantitative importance of this effect we consider
$\gamma a$ conversion in the Galaxy. We consider propagation in the $x$
direction in a transverse $B$-field, leading to
\begin{equation}
-i\partial_x\begin{pmatrix}A_\parallel\cr a\cr\end{pmatrix}
=\begin{pmatrix}\chi_{\rm tot}\omega&g_{a\gamma}B/2\cr
                g_{a\gamma}B/2& -m_a^2/2\omega\cr\end{pmatrix}
\begin{pmatrix}A_\parallel\cr a\cr\end{pmatrix}\,.
\end{equation}
Here, $\chi_{\rm tot}=\chi_{\rm CMB}+\chi_B$ with the CMB and $B$-field
contributions. In addition, there can be contributions from other
photon populations. The oscillation probability (distance $L$) is
\begin{equation}
P_{\gamma\to a}=(\Delta_{a\gamma}L)^2\,
\frac{\sin^2(\Delta_{\rm osc}L/2)}{(\Delta_{\rm osc}L/2)^2}\,,
\end{equation}
where $\Delta_{a\gamma}=g_{a\gamma}B/2$ and the ``oscillation wave number''
is  $\Delta_{\rm osc}=[(\chi_{\rm EM}\omega+m_a^2/2\omega)^2 +
(g_{a\gamma}B)^2]^{1/2}$.

In the Galaxy, magnitudes of the matrix components in typical
scenarios are\footnote{In the expression for $\chi_B$ we now include
the factor 21/11 for $A_\parallel$.}
\begin{subequations}
\begin{eqnarray}
g_{a\gamma}B/2&=&+1.52\times10^{-2}~{\rm kpc}^{-1}~g_{11}B_{\mu{\rm G}}\,,\\
\chi_{\rm CMB}\omega&=&+0.80\times10^{-4}~{\rm kpc}^{-1}~\omega_{\rm TeV}\,,\\
\chi_B\omega&=&+1.44\times10^{-5}~{\rm kpc}^{-1}~B_{\mu{\rm G}}^2\omega_{\rm TeV}\,,\\
-m_a^2/2\omega&=&-0.78\times10^{-4}~{\rm kpc}^{-1}~m_{\rm neV}^2/\omega_{\rm TeV}\,,\\
-\omega_{\rm pl}^2/2\omega&=&-1.08\times10^{-10}~{\rm kpc}^{-1}~n_{3}/\omega_{\rm TeV}\,,
\end{eqnarray}
\end{subequations}
where $g_{11}=g_{a\gamma}/(10^{-11}~{\rm GeV}^{-1})$, $B_{\mu{\rm G}} =
B/(1~\mu{\rm G})$, $\omega_{\rm TeV}=\omega/(1~{\rm TeV})$, $m_{\rm
neV}=m_a/(10^{-9}~{\rm eV})$ and $n_3=n_e/(10^{-3}~{\rm cm}^{-3})$. For
completeness we have included the electron contribution, which is completely
negligible.

The term $g_{a\gamma}B/2$ exceeds all others, corresponding to maximal
mixing. Indeed, the considered ALP masses are in the neV range to achieve
this effect. The ALP and photon dispersion relations have opposite sign
(timelike vs.\ spacelike) so that the two effects add up in the expression
for $\Delta_{\rm osc}$. They cannot cancel each other and must be separately
small to achieve large mixing.

Therefore, $\gamma\gamma$ dispersion will be unimportant only if $\chi_{\rm
EM}\omega\ll g_{a\gamma}B$. The above parameters satisfy this condition, but
maximal mixing can be lost, and $\gamma\gamma$ dispersion becomes important,
for smaller $g_{a\gamma}$ or weaker $B$ (e.g.\ in intergalactic space).
Likewise, $\omega\agt 100$~TeV, keeping all else fixed, implies $\chi_{\rm
CMB}\omega\sim g_{a\gamma}B$.

Moreover, the radiation fields in the Galaxy, in the TeV source regions, and
the EBL provide additional contributions. However, these photons typically
exceed the pair-production threshold so that we need to go beyond the
low-energy limit to estimate their dispersive effect on TeV $\gamma$
propagation.

\section{Beyond Euler-Heisenberg}

So far, our results apply when pair creation can be neglected. In a static
$B$-field, this is true when the dynamical parameter $eB
\omega/2m_e^3=2.21\times10^{-14}\,B_{\mu{\rm G}}\omega_{\rm TeV}\ll 1$
\cite{Adler:1971wn, Tsai:1975iz, Karbstein:2013ufa}. This condition is easily
fulfilled in our context.

Pair production in the CMB becomes important for $\omega\agt 100$~TeV.
However, for TeV $\gamma$ rays propagating in the EBL or the galactic star
light, the Euler-Heisenberg limit breaks down. The only pertinent literature
is John Toll's often-cited PhD Thesis (1952) \cite{Toll:1952rq} which we have
actually found in our library. However, his results may be
incorrect\footnote{Toll writes that he has solved the Kramers-Kronig
  integral analytically (his Eq.~2.2-10), but the result on p.~54 fails to
  define his quantity $b$. The plot of $d\sigma/d\Omega|_{\rm
  forward}$ in Fig.~2.2~B, which is proportional to the squared
  forward-scattering amplitude and thus to $|1-n_{\rm refr}|^2$, does not have
  a zero. Our numerical integration of
  his equations does not reproduce his Fig.~2.2~B.}
and we perform our own analysis.

\begin{figure}[b]
\includegraphics[width=0.85\columnwidth]{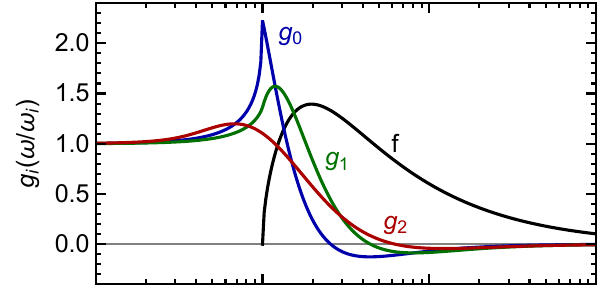}
\vskip2pt
\includegraphics[width=0.85\columnwidth]{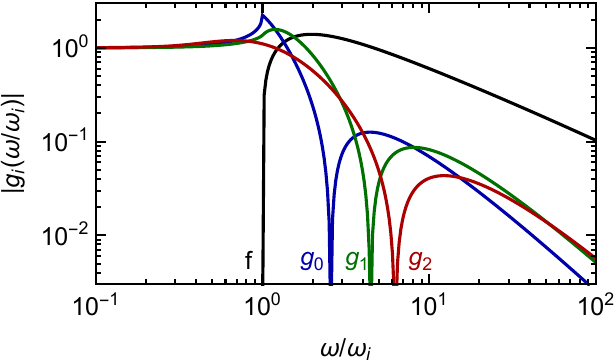}
\caption{Photon-photon dispersion as a function of energy.
{\it Black line:} Absorptive part $f(\omega/\omega_0)$
defined in Eq.~(\ref{eq:fabs}).
{\it Blue line:} Dispersive part for background photons with fixed energy and
fixed direction, $g_0(\omega/\omega_0)$, defined in Eq.~(\ref{eq:gdef}).
{\it Green line:} Angle average for isotropic distribution with fixed energy
$\omega_{\rm B}$, $g_1(\omega/\omega_1)$, defined in Eq.~(\ref{eq:g1def}).
{\it Red line:} Thermal, isotropic average, $g_2(\omega/\omega_2)$,
defined in Eq.~(\ref{eq:g2def}).}
\label{fig:gplot}
\end{figure}

The dispersive part $\chi$ of the refractive index is related to the
imaginary part $\kappa=\Gamma/2\omega$ (absorption rate $\Gamma$) by the
Kramers-Kronig relation~\cite{Toll:1952rq, Toll:1956cya, Jackson:1975}
\begin{equation}
\chi(\omega)=\frac{1}{\pi}~\dashint_0^\infty d\omega'\,
\frac{\Gamma(\omega')}{\omega'^2-\omega^2}\,,
\end{equation}
where the integral denotes the Cauchy principal value. We assume background
photons with number density $n_{\rm B}$, energy $\omega_{\rm B}$, and
direction $\theta$ relative to the test photon. Then $\Gamma=(1-\cos\theta)
n_{\rm B}\sigma_{\gamma\gamma}$, where
$\sigma_{\gamma\gamma}=(\pi\alpha^2/2m_e^2)\,f(u)$ is the total
$\gamma\gamma\to e^+e^-$ cross section \cite{Breit:1934zz, Karplus:1950zz,
Jauch:1955}. Here $u=\omega/\omega_0$ and $\omega_0$ is the threshold energy
for pair production defined by $\omega_0\omega_{\rm B}(1-\cos\theta)=2m_e^2$
and
\begin{eqnarray}\label{eq:fabs}
f(u)&=&\frac{4u(u+1)-2}{u^3}\log\left(\sqrt{u}+\sqrt{u-1}\right)
\nonumber\\
&&{}-\frac{2(u+1)\sqrt{u-1}}{u^{5/2}}\,.
\end{eqnarray}
We show this function in Fig.~\ref{fig:gplot} as a black line.
The dispersive part of the refractive index therefore is
\begin{equation}\label{eq:chigeneral}
\chi(\omega)=\chi_{\rm EM}\,\frac{3(1-\cos\theta)^2}{4}\,g_0(\omega/\omega_0)\,,
\end{equation}
where $\chi_{\rm EM}$ is given in Eq.~(\ref{eq:chiEM}) with
$\rho_{\rm EM}=n_{\rm B}\omega_{\rm B}$ and
\begin{equation}\label{eq:gdef}
g_0(u)=\frac{45}{44}~\dashint_1^\infty du'\,
\frac{f(u')}{u'^2-u^2}\,.
\end{equation}
In the limit $u=0$ we find $g_0(u)=1$. Taking the angle average
in Eq.~(\ref{eq:chigeneral}) we find
$\frac{1}{2}\int_{-1}^{+1}(1-\cos\theta)^2d\cos\theta=\frac{4}{3}$,
i.e., the Kramers-Kronig transformation reproduces the
Euler-Heisenberg result stated earlier.

We note that background photons collinear with the test photon produce no
refractive effect, head-on moving ones produce the largest effect. This
behavior is similar to neutrino-neutrino refraction, although for neutrinos
the factor is only $1-\cos\theta$. Of course, the origin of the angle
dependence is very different in the two cases.

We show $g_0(\omega/\omega_0)$ in Fig.~\ref{fig:gplot} as a blue line
and note that it has a
cusp at the pair-creation threshold $\omega=\omega_0$, it crosses zero at
$\omega=2.5661\,\omega_0$, and then approaches zero asymptotically from
below. Unsurprisingly, the dispersion relation becomes timelike for high
energies.

However, it does not approach the form of an effective mass, i.e., $g_0(u)$
does not become proportional to $-1/u^2$ for high energies. Note that
$f(u)\propto u^{-1}\log(u)$ for large $u$ and the integral $\int du' f(u')$
does not converge. In this context we note that for very large photon
energies, the process $\gamma\gamma\to e^-e^+e^-e^+$ acquires a constant
cross section of order $\alpha^4/m_e^2$ and thus becomes more important than
$\gamma\gamma\to e^-e^+$ \cite{Dittrich:2000zu, Cheng:1970ef}. However, this
very high-energy behavior is not important in our context.

Next we consider an isotropic distribution of background photons with fixed
energy $\omega_{\rm B}$. As a fiducial energy we now use the pair-creation
threshold $\omega_1=m_e^2/\omega_{\rm B}$ for a head-on moving photon. The
refractive index is then $\chi(\omega)=\chi_{\rm EM}\,g_1(\omega/\omega_1)$,
where
\begin{equation}\label{eq:g1def}
g_1(v)=\frac{3}{8}\int_{-1}^{+1}d\mu\,
(1-\mu)^2\,g_0\left(v\,\frac{1-\mu}{2}\right)
\end{equation}
is the angle average with $\mu=\cos\theta$. We plot $g_1(v)$ in
Fig.~\ref{fig:gplot} as a green line.

Finally we consider a grey-body photon distribution with fixed energy density
$\rho_{\rm EM}$ and thermal spectrum. As a fiducial energy we use
$\omega_2=m_e^2/\langle \omega_{\rm B}\rangle$ in terms of the average
$\langle\omega_{\rm B}\rangle=(\pi^4/30\zeta_3)\,T\approx2.701\,T$. The
refractive index is then $\chi(\omega)=\chi_{\rm EM}\,g_2(\omega/\omega_2)$,
where\footnote{In earlier versions of this Eprint and in the version published
in PRD we had taken the average over the number distribution of background
photons rather than their energy distribution, i.e., we had used
$x^2$ in the integrand instead of $x^3$ and concomitant normalization
factor $1/(2\zeta_3)$ instead of $15/\pi^4$. This correction has also changed
the red line in Fig.~1 and the zero intercept and minimum given after
Eq.~(\ref{eq:g2def}). Figures~2 and 3 have also been updated, where only the
CMB curves were affected by this oversight. There is no change of the overall results of this paper.
We thank Hendrik Vogel and Ranjan Laha for spotting this issue.}
\begin{equation}\label{eq:g2def}
g_2(w)=\frac{15}{\pi^4}\int_0^\infty dx\,\frac{x^3}{e^x-1}\,
g_1\left(w\,x\,\frac{30\zeta_3}{\pi^4}\right)
\end{equation}
is the thermal average in terms of $x=\omega_{\rm B}/T$. We plot
$g_2(w)$ in Fig.~\ref{fig:gplot} as a red line. It crosses to negative values
at approximately $\omega=6.25\,\omega_2=2.31\,m_e^2/T$ and reaches its minimum of $-0.043$
at $\omega=12.3\,\omega_2=4.55\,m_e^2/T$.

We conclude that for a given $\rho_{\rm EM}$, low-energy background photons
below the pair threshold are most important for dispersion, whereas photons
around the pair threshold are most important for absorption. In intergalactic
space, the EBL is the most important photon population for the absorption of
TeV $\gamma$ rays, but $\rho_{\rm EBL}$ is only about $0.10\,\rho_{\rm CMB}$.
For TeV photons, the CMB is far more important for dispersion than the EBL.

\section{Local radiation density}

\begin{figure}[b]
\includegraphics[width=0.85\columnwidth]{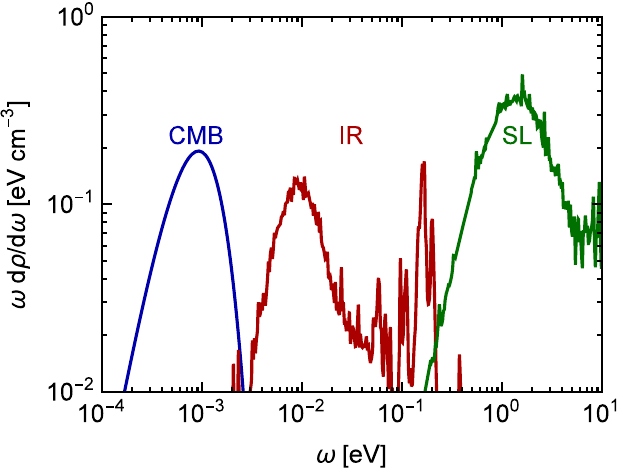}
\caption{Interstellar radiation field in the Galaxy near the Sun
\cite{Moskalenko:2005ng, Porter:2008ve}, consisting of the CMB, infrared
radiation (IR) and star light (SL). (Extracted from the
GALPROP code \cite{Vladimirov:2010aq} and available in Ref.~\cite{Buch:2014}.)}
\label{fig:ISRF}
\vskip12pt
\includegraphics[width=0.85\columnwidth]{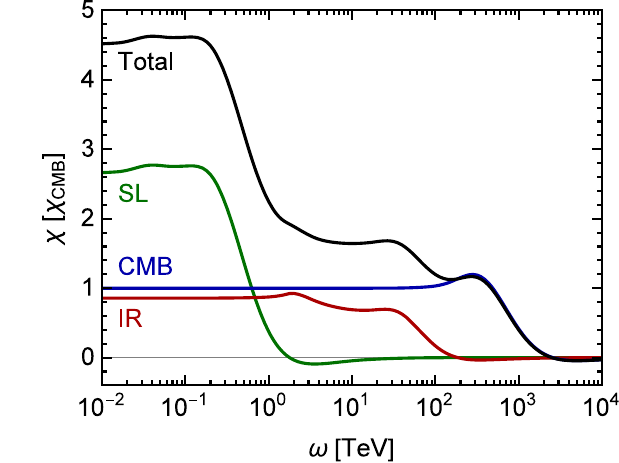}
\caption{Photon-photon dispersion in the solar neighborhood based
on the EM radiation field components shown in Fig.~\ref{fig:ISRF}.}
\label{fig:refr}
\end{figure}

The radiation density in the Galaxy can far exceed the CMB. The main
component is star light (SL) which, however, is partly processed by dust to
form infrared radiation (IR). In Fig.~\ref{fig:ISRF} we show the estimated
spectral energy distribution of these components in the Galaxy near the solar
neighborhood \cite{Moskalenko:2005ng, Porter:2008ve, Vladimirov:2010aq}. The
IR energy density is comparable to the CMB whereas the SL provides about 2.6
times more energy. At smaller galactocentric distances, the non-CMB
contributions are much larger.

Another way of estimating the importance of star light is to use the total
galactic luminosity of about $5\times10^{10}\,L_\odot$ and, if the source
were concentrated at the galactic center, would provide $\rho_{\rm
EM}/\rho_{\rm CMB}\sim(12~{\rm kpc}/r)^2$. Of course, the disk
geometry
requires a detailed model, e.g., the one of the
GALPROP code \cite{Vladimirov:2010aq} that we used for
Fig.~\ref{fig:ISRF}.

The corresponding $\gamma\gamma$ refractive index is shown in
Fig.~\ref{fig:refr} as a function of the test-photon energy. For $\omega\alt
200$~GeV, all background radiations contribute essentially with their
Euler-Heisenberg strength, whereas for higher energies, first the star light
and then the infrared radiation drop out. The CMB contribution becomes small
and finally negative only at $\omega\agt2000$~TeV.

\section{Other effects}

Photon-photon refraction leads to deflection, e.g., in the radiation field of
the Sun. In the Euler-Heisenberg limit and for photons grazing the Sun, we
find an energy-independent deflection of $6.7\times10^{-24}$ arcsec, much
smaller than the gravitational deflection of $1.75$ arcsec. Photon-photon
dispersion matters only in the context of $\gamma$-ALP oscillations where
interference with the ALP dispersion enhances the effect

In the early universe, there is a brief epoch when $\gamma\gamma$ dispersion
dominates. As the universe cools, the $e^+e^-$ density is
$n_{e^-e^+}=2^{1/2}(m_eT/\pi)^{3/2}e^{-m_e/T}$, producing $\omega_{\rm
pl}^2=4\pi\alpha n_{e^-e^+}/m_e=6.08\times10^9~{\rm eV}^2 (T/m_e)^{3/2}
e^{-m_e/T}$. Photons provide $\chi_{\rm EM}=1.676\times10^{-16}\,T_{\rm
keV}^4$, corresponding to $m_{\rm eff}^2=-2\chi_{\rm EM}\omega^2$ and a
thermal average $\langle m_{\rm eff}^2\rangle\sim-3.47\times10^{-9}~{\rm
eV}^2\,T_{\rm keV}^6$. This is similar to $-\omega_{\rm pl}^2$ at $T=30~{\rm
keV}$, in agreement with the crossover shown in Fig.~3.6 of
Ref.~\cite{Dittrich:2000zu}. The cosmic $e/\gamma$ ratio is about $5.3\times
10^{-10}$ so that $\omega_{\rm pl}^2=2.32\times10^{-8}~{\rm eV}^2~T_{\rm
keV}^3$. It takes over from $\gamma\gamma$ dispersion at $T\sim 2~{\rm keV}$.

Therefore, in the primordial plasma, $\gamma\gamma$ dispersion dominates when
$2~{\rm keV}\alt T\alt 30~{\rm keV}$, providing photons with a spacelike
dispersion relation. Note, however, that the photon gas does not support
longitudinal excitations and does not contribute to Debye
screening~\cite{Thoma:2000fd}.

We also mention a recent study of the impact of photon-photon interaction on
the polarization of CMB photons after recombination~\cite{Sawyer:2014maa},
although the effect looks extremely small. Photon-photon interaction is a
polarization-dependent effect and therefore can lead to nontrivial
birefringence effects~\cite{Kotkin:1996nf, Sawyer:2004nj}.

\section{Conclusions}

A photon gas is a dispersive medium for photon propagation. The ubiquitous
CMB alone produces $n_{\rm refr}=1+0.511\times10^{-42}$, independently of energy
if $\omega\alt 1000$~TeV. This tiny effect dominates the dispersion of TeV
gamma rays and, while it has always been ignored, can modify the oscillation
between TeV gamma rays and axionlike particles in astrophysical magnetic
fields.

If the energies of the background photons exceed the pair-creation threshold,
the dispersion effect decreases, i.e., soft background photons are more
important. Therefore, even though radiation in the Galaxy or the source
regions can far exceed the CMB, their harder spectra prevent them from having
a large impact on dispersion except for relatively small energies of
$\omega\alt100$~GeV. On the other hand, $\gamma\gamma$ dispersion is weaker
for smaller $\omega$, so while the relative importance of local radiation
fields is larger for smaller $\omega$, the absolute importance of the overall
effect decreases.

Photon-ALP oscillations depend on the $a\gamma$ interaction strength, the
$B$-field strength and spatial distribution, the ALP mass, the photon energy,
and, as a new ingredient, the density and spectrum of background photons. It
is fortuitous that for many scenarios considered in the literature,
$\gamma\gamma$ dispersion will be a benign effect and does not exclude that
ALPs could be important for TeV gamma propagation in the universe. On the
other hand, the effect is large enough that it cannot be summarily
dismissed---its quantitative importance has to be evaluated in every
individual case.

TeV gamma rays propagating in the universe provide an intriguing example
where $\gamma\gamma$ dispersion, despite its intrinsic weakness, can be of
practical interest.

{\ }

{\ }

\section*{ACKNOWLEDGMENTS}

We thank H.~Gies, F.~Karbstein, A.~Mirizzi, D.\ Pane\-que, M.~Thoma and
G.~Sigl for pointing us to some of the pertinent literature and M.~Cirelli,
G.~Giesen and M.~Taoso for help with the data for the interstellar radiation
field. We acknowledge partial support by the Michail-Lomonosov-Program of the
German Academic Exchange Service (DAAD) and the Ministry of Education and
Science of the Russian Federation (Project No.~11.9164.2014), the Russian
Foundation for Basic Research (Project No.\ 15-02-06033-a), the Dynasty
Foundation, the Deutsche Forschungsgemeinschaft (DFG) under Grant No.\
EXC-153 (Excellence Cluster ``Universe''), and the Research Executive Agency
(REA) of the European Union under Grant No.\ PITN-GA-2011-289442 (FP7 Initial
Training Network ``Invisibles'').

{\ }



\begin{thebibliography}{00}

\bibitem{Aharonian:2004yt}
  F.~A.~Aharonian,
  {\it Very high energy cosmic gamma radiation}
  (River Edge, USA: World Scientific, 2004).

\bibitem{Aharonian:2013}
  F.~A.~Aharonian, L.~Bergstr\"om and C.~Dermer,
  {\it Astrophysics at Very High Energies\/}
  (Saas-Fee Advanced Course 40. Swiss Society for Astrophysics and
  Astronomy, 2013).

\bibitem{Hinton:2009zz}
  J.~A.~Hinton and W.~Hofmann,
  Annu.\ Rev.\ Astron.\ Astrophys.\  {\bf 47}, 523 (2009).

\bibitem{Lorenz:2012nw}
  E.~Lorenz and R.~Wagner,
  Eur.\ Phys.\ J.\ H {\bf 37}, 459 (2012).

\bibitem{Paneque:2012kv}
  D.~Paneque,
  J.\ Phys.\ Conf.\ Ser.\  {\bf 375}, 052020 (2012).

\bibitem{Stecker:1992wi}
  F.~W.~Stecker, O.~C.~De Jager and M.~H.~Salamon,
  Astrophys.\ J.\  {\bf 390}, L49 (1992).

\bibitem{Dwek:2012nb}
  E.~Dwek and F.~Krennrich,
  Astropart.\ Phys.\  {\bf 43}, 112 (2013).

\bibitem{Biteau:2015xpa}
  J.~Biteau and D.~A.~Williams,
  Astrophys.\ J.\  {\bf 812}, 60 (2015).

\bibitem{Stecker:2001vb}
  F.~W.~Stecker and S.~L.~Glashow,
  Astropart.\ Phys.\  {\bf 16}, 97 (2001).

\bibitem{Liberati:2009pf}
  S.~Liberati and L.~Maccione,
  Annu.\ Rev.\ Nucl.\ Part.\ Sci.\ {\bf 59}, 245 (2009).

\bibitem{Liberati:2013xla}
  S.~Liberati,
  Class.\ Quant.\ Grav.\  {\bf 30}, 133001 (2013).

\bibitem{Kuster:2008zz}
  M.~Kuster, G.~Raffelt and B.~Beltr\'an (eds.),
  {\it Axions: Theory, cosmology, and experimental searches},
  Lect.\ Notes Phys.\  {\bf 741}, 1 (2008).

\bibitem{Jaeckel:2010ni}
  J.~Jaeckel and A.~Ringwald,
  Annu.\ Rev.\ Nucl.\ Part.\ Sci.\ {\bf 60}, 405 (2010).

\bibitem{Hewett:2012ns}
  J.~L.~Hewett {\it et al.},
  {\em Fundamental Physics at the Intensity Frontier},
  arXiv:1205.2671.

\bibitem{Agashe:2014kda}
  A.~Ringwald, L.~J.~Rosenberg and G.~Rybka,
  ``Axions and other similar particles,'' in:
  K.~A.~Olive {\it et al.}  (Particle Data Group),
  Chin.\ Phys.\ C {\bf 38}, 090001 (2014).

\bibitem{Sikivie:1983ip}
  P.~Sikivie,
  Phys.\ Rev.\ Lett.\  {\bf 51}, 1415 (1983);
  Erratum {\it ibid.}\ {\bf 52}, 695 (1984).

\bibitem{Raffelt:1987im}
  G.~Raffelt and L.~Stodolsky,
  Phys.\ Rev.\ D {\bf 37}, 1237 (1988).

\bibitem{Csaki:2003ef}
  C.~Cs\'aki, N.~Kaloper, M.~Peloso and J.~Terning,
  JCAP {\bf 0305}, 005 (2003).

\bibitem{Mirizzi:2007hr}
  A.~Mirizzi, G.~G.~Raffelt and P.~D.~Serpico,
  Phys.\ Rev.\ D {\bf 76}, 023001 (2007).

\bibitem{Hooper:2007bq}
  D.~Hooper and P.~D.~Serpico,
  Phys.\ Rev.\ Lett.\ {\bf 99}, 231102 (2007).

\bibitem{De Angelis:2007yu}
  A.~De Angelis, O.~Mansutti and M.~Roncadelli,
  Phys.\ Lett.\ B {\bf 659}, 847 (2008).

\bibitem{De Angelis:2007dy}
  A.~De Angelis, M.~Roncadelli and O.~Mansutti,
  Phys.\ Rev.\ D {\bf 76}, 121301 (2007).

\bibitem{Hochmuth:2007hk}
  K.~A.~Hochmuth and G.~Sigl,
  Phys.\ Rev.\ D {\bf 76}, 123011 (2007),

\bibitem{Simet:2007sa}
  M.~Simet, D.~Hooper and P.~D.~Serpico,
  Phys.\ Rev.\ D {\bf 77}, 063001 (2008).

\bibitem{DeAngelis:2008sk}
  A.~De Angelis, O.~Mansutti, M.~Persic and M.~Roncadelli,
  Mon.\ Not.\ Roy.\ Astron.\ Soc.\ {\bf 394}, L21 (2009).

\bibitem{Fairbairn:2009zi}
  M.~Fairbairn, T.~Rashba and S.~V.~Troitsky,
  Phys.\ Rev.\ D {\bf 84}, 125019 (2011).

\bibitem{SanchezConde:2009wu}
  M.~A.~S\'anchez-Conde, D.~Paneque, E.~Bloom, F.~Prada and A.~Dom{\'\i}nguez,
  Phys.\ Rev.\ D {\bf 79}, 123511 (2009).

\bibitem{Bassan:2009gy}
  N.~Bassan and M.~Roncadelli,
  arXiv:0905.3752.

\bibitem{Mirizzi:2009aj}
  A.~Mirizzi and D.~Montanino,
  JCAP {\bf 0912}, 004 (2009).

\bibitem{Dominguez:2011xy}
  A.~Dom{\'\i}nguez, M.~A.~S\'anchez-Conde and F.~Prada,
  JCAP {\bf 1111}, 020 (2011).

\bibitem{DeAngelis:2011id}
  A.~De Angelis, G.~Galanti and M.~Roncadelli,
  Phys.\ Rev.\ D {\bf 84}, 105030 (2011).

\bibitem{Tavecchio:2012um}
  F.~Tavecchio, M.~Roncadelli, G.~Galanti and G.~Bonnoli,
  Phys.\ Rev.\ D {\bf 86}, 085036 (2012).

\bibitem{Wouters:2012qd}
  D.~Wouters and P.~Brun,
  Phys.\ Rev.\ D {\bf 86}, 043005 (2012).

\bibitem{Horns:2012kw}
  D.~Horns, L.~Maccione, M.~Meyer, A.~Mirizzi, D.~Montanino and M.~Roncadelli,
  Phys.\ Rev.\ D {\bf 86}, 075024 (2012).

\bibitem{Mena:2013baa}
  O.~Mena and S.~Razzaque,
  JCAP {\bf 1311}, 023 (2013).

\bibitem{Wouters:2013eka}
  D.~Wouters and P.~Brun,
  JCAP {\bf 1401}, 016 (2014).

\bibitem{Meyer:2013pny}
  M.~Meyer, D.~Horns and M.~Raue,
  Phys.\ Rev.\ D {\bf 87}, 035027 (2013).

\bibitem{Meyer:2014epa}
  M.~Meyer, D.~Montanino and J.~Conrad,
  JCAP {\bf 1409}, 003 (2014).


\bibitem{Tavecchio:2014yoa}
  F.~Tavecchio, M.~Roncadelli and G.~Galanti,
  Phys.\ Lett.\ B {\bf 744}, 375 (2015).
  
\bibitem{Reesman:2014ova}
  R.~Reesman and T.~P.~Walker,
  JCAP {\bf 1408}, 021 (2014).

\bibitem{Harris:2014nza}
  J.~Harris and P.~M.~Chadwick,
  JCAP {\bf 1410}, 018 (2014).

\bibitem{Meyer:2014gta}
  M.~Meyer and J.~Conrad,
  JCAP {\bf 1412}, 016 (2014).

\bibitem{Galanti:2015rda}
  G.~Galanti, M.~Roncadelli, A.~De Angelis and G.~F.~Bignami,
  arXiv:1503.04436.

\bibitem{Dasgupta:2010ck}
  B.~Dasgupta and G.~G.~Raffelt,
  Phys.\ Rev.\ D {\bf 82}, 123003 (2010).

\bibitem{Tarrach:1983cb}
  R.~Tarrach,
  Phys.\ Lett.\ B {\bf 133}, 259 (1983).

\bibitem{Barton:1989dq}
  G.~Barton,
  Phys.\ Lett.\ B {\bf 237}, 559 (1990).

\bibitem{Latorre:1994cv}
  J.~I.~Latorre, P.~Pascual and R.~Tarrach,
  Nucl.\ Phys.\ B {\bf 437}, 60 (1995).

\bibitem{Kong:1998ic}
  X.-W.~Kong and F.~Ravndal,
  Nucl.\ Phys.\ B {\bf 526}, 627 (1998).

\bibitem{Thoma:2000fd}
  M.~H.~Thoma,
  Europhys.\ Lett.\  {\bf 52}, 498 (2000).

\bibitem{Dittrich:2000zu}
  W.~Dittrich and H.~Gies,
  {\em Probing the quantum vacuum},
  Springer Tracts Mod.\ Phys.\  {\bf 166}, 1 (2000).

\bibitem{Toll:1952rq}
  J.~S.~Toll,
  {\it The dispersion relation for light and its application
  to problems involving electron pairs\/}
  (PhD Thesis, Princeton University, 1952).

\bibitem{Adler:1971wn}
  S.~L.~Adler,
  Ann.\ Phys.\ (N.Y.) {\bf 67}, 599 (1971).

\bibitem{Tsai:1975iz}
  W.-Y.~Tsai and T.~Erber,
  Phys.\ Rev.\ D {\bf 12}, 1132 (1975).

\bibitem{Karbstein:2013ufa}
  F.~Karbstein,
  Phys.\ Rev.\ D {\bf 88}, 085033 (2013).

\bibitem{Toll:1956cya}
  J.~S.~Toll,
  Phys.\ Rev.\ {\bf 104}, 1760 (1956).

\bibitem{Jackson:1975}
  J.~D.~Jackson,
  {\em Classical Electrodynamics, 2nd ed.}
  (Wiley, New York, 1975).

\bibitem{Breit:1934zz}
  G.~Breit and J.~A.~Wheeler,
  Phys.\ Rev.\  {\bf 46}, 1087 (1934).

\bibitem{Karplus:1950zz}
  R.~Karplus and M.~Neuman,
  Phys.\ Rev.\  {\bf 83}, 776 (1951).

\bibitem{Jauch:1955}
  J.~M.~Jauch and F.~Rohrlich,
  {\em The theory of photons and electrons\/}
  (Addison-Wesley, Cambridge, MA, 1955).

\bibitem{Cheng:1970ef}
  H.~Cheng and T.~T.~Wu,
  Phys.\ Rev.\ D {\bf 1}, 3414 (1970).

\bibitem{Moskalenko:2005ng}
  I.~V.~Moskalenko, T.~A.~Porter and A.~W.~Strong,
  Astrophys.\ J.\  {\bf 640}, L155 (2006).

\bibitem{Porter:2008ve}
  T.~A.~Porter, I.~V.~Moskalenko, A.~W.~Strong, E.~Orlando and L.~Bouchet,
  Astrophys.\ J.\  {\bf 682}, 400 (2008).

\bibitem{Vladimirov:2010aq}
  A.~E.~Vladimirov, S.~W.~Digel, G.~J\'ohannesson, P.~F.\ Michelson, I.~V.~Moskalenko, P.~L.~Nolan, E.~Orlando, T.~A.~Porter and A.~W.~Strong,
  Comput.\ Phys.\ Commun.\  {\bf 182}, 1156 (2011).
  See also the GALPROP web page at
  http://galprop.stanford.edu/

\bibitem{Buch:2014}
  J.~Buch, M.~Cirelli, G.~Giesen, and M.~Taoso,
  work in progress (2015).

\bibitem{Sawyer:2014maa}
  R.~F.~Sawyer,
  arXiv:1408.5434.

\bibitem{Kotkin:1996nf}
  G.~L.~Kotkin and V.~G.~Serbo,
  Phys.\ Lett.\ B {\bf 413}, 122 (1997).

\bibitem{Sawyer:2004nj}
  R.~F.~Sawyer,
  Phys.\ Rev.\ Lett.\  {\bf 93}, 133601 (2004).

\end{thebibliography}
\end{document}